\documentclass[aps,prl,twocolumn,showpacs,superscriptaddress,reprint]{revtex4-1}

\usepackage{amsmath,amssymb,bm,graphicx}
\usepackage[colorlinks=true,urlcolor=blue,linkcolor=blue,citecolor=blue,bookmarks=false]{hyperref}

\begin{document}

\newcommand{\cso}{Cu$_{2}$OSeO$_{3}$}
\newcommand{\hc}{$H_{\text{c2}}$}

\title{Low spin wave damping in the insulating chiral magnet \cso}

\author{I.~Stasinopoulos}
\affiliation{Physik Department E10, Technische Universit\"at M\"unchen, D-85748 Garching, Germany}

\author{S.~Weichselbaumer}
\affiliation{Physik Department E10, Technische Universit\"at M\"unchen, D-85748 Garching, Germany}

\author{A.~Bauer}
\affiliation{Physik Department E51, Technische Universit\"at M\"unchen, D-85748 Garching, Germany}

\author{J.~Waizner}
\affiliation{Institute for Theoretical Physics, Universit\"at zu K\"oln, D-50937 K\"oln, Germany}

\author{H.~Berger}
\affiliation{Institut de Physique de la Mati\`ere Complexe, \'Ecole Polytechnique F\'ed\'erale de Lausanne, 1015 Lausanne, Switzerland}

\author{S.~Maendl}
\affiliation{Physik Department E10, Technische Universit\"at M\"unchen, D-85748 Garching, Germany}

\author{M.~Garst}
\affiliation{Institute for Theoretical Physics, Universit\"at zu K\"oln, D-50937 K\"oln, Germany}
\affiliation{Institut f\"ur Theoretische Physik, Technische Universit\"at Dresden, D-01062 Dresden, Germany}

\author{C.~Pfleiderer}
\affiliation{Physik Department E51, Technische Universit\"at M\"unchen, D-85748 Garching, Germany}

\author{D.~Grundler}\thanks{Electronic mail: dirk.grundler@epfl.ch}
\affiliation{Institute of Materials and Laboratory of Nanoscale Magnetic Materials and Magnonics (LMGN), \'Ecole Polytechnique F\'ed\'erale de Lausanne (EPFL), Station 12, 1015 Lausanne, Switzerland}

\date{\today}

\begin{abstract}
Chiral magnets with topologically nontrivial spin order such as Skyrmions have generated enormous interest in both fundamental and applied sciences. We report broadband microwave spectroscopy performed on the insulating chiral ferrimagnet \cso. For the damping of magnetization dynamics we find a remarkably small Gilbert damping parameter of about $1\times10^{-4}$ at 5\,K. This value is only a factor of 4 larger than the one reported for the best insulating ferrimagnet yttrium iron garnet. We detect a series of sharp resonances and attribute them to confined spin waves in the mm-sized samples. Considering the small damping, insulating chiral magnets turn out to be promising candidates when exploring non-collinear spin structures for high frequency applications.\end{abstract}

\pacs{76.50.+g, 74.25.Ha, 4.40.Az, 41.20.Jb}%
\keywords{Skyrmions; spin dynamics; Gilbert-damping; magnonics; coplanar waveguides; chiral magnets}
\maketitle

\begin{figure}[t!]
	\includegraphics[width=1.0\linewidth]{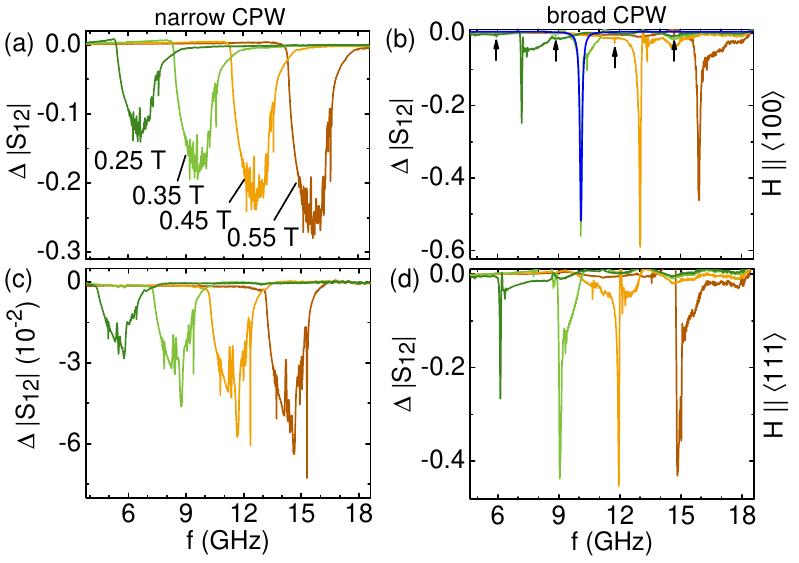}
	\caption{(Color online) Spectra $\Delta |S_{\text{12}}|$ obtained at T = 5\,K for different $\mathbf{H}$ using (a) a narrow and (b) broad CPW when $\mathbf{H}||\langle100\rangle$ on sample S1. Corresponding spectra taken on sample S2 for $\mathbf{H}||\langle111\rangle$ are shown in (c) and (d), respectively. Note the strong and sharp resonances in (b) and (d) when using the broad CPW that provides a much more homogeneous excitation field $\mathbf{h}$. Arrows mark resonances that have a field-independent offset with the corresponding main peaks and are attributed to standing spin waves. An exemplary Lorentz fit curve is shown in blue color in (b).}
	\label{spectra}
\end{figure}

\begin{figure}[t!]
	\includegraphics[width=1.0\linewidth]{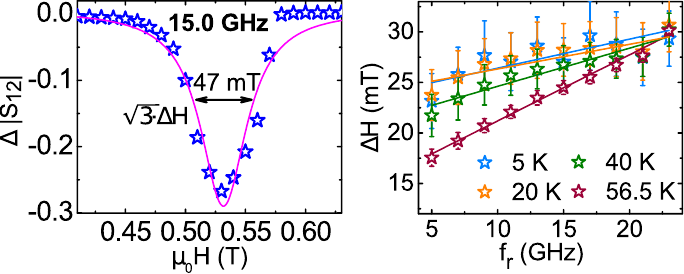}
	\caption{(Color online) (a) Lorentz curve (magenta line) fitted to a resonance (symbols) measured at $f=15$ GHz as a function of $H$ at 5 K. (b) Frequency dependencies of linewidths $\Delta H$ (symbols) for four different $T$. We performed the $\sqrt{3}$-correction.  The slopes of linear fits (straight lines) following Eq.\,\ref{DeltaH} are considered to reflect the intrinsic damping parameters $\alpha_{\text{intr}}$.}
	\label{deltaH_narrowCPW}
\end{figure}

\begin{figure}[h!t!]
	\includegraphics[width=1.0\linewidth]{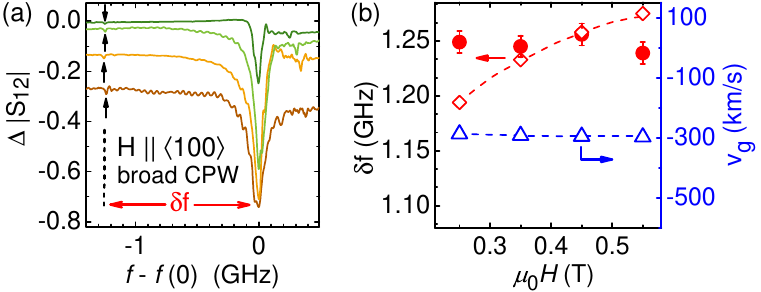}
	\caption{(Color online) (a) Spectra of Fig.\,\ref{spectra} (b) replotted as $f-f_{\rm FMR}(H)$ for different $H$ such that all main peaks are at zero frequency and the field-independent frequency splitting $\delta f$ becomes visible. The numerous oscillations seen particularly on the bottom most curve are artefacts from the calibration routine. (b) Experimentally evaluated (filled circles) and theoretically predicted (diamonds) splitting $\delta f$ using dispersion relations for a platelet. Calculated group velocity $v_{\rm g}$ at $k=\pi/(0.3~{\rm mm})$. Dashed lines are guides to the eyes.}
	\label{Iguchi_vs_exp}
\end{figure}

\begin{figure}[h!t!]
	\includegraphics[width=1.0\linewidth]{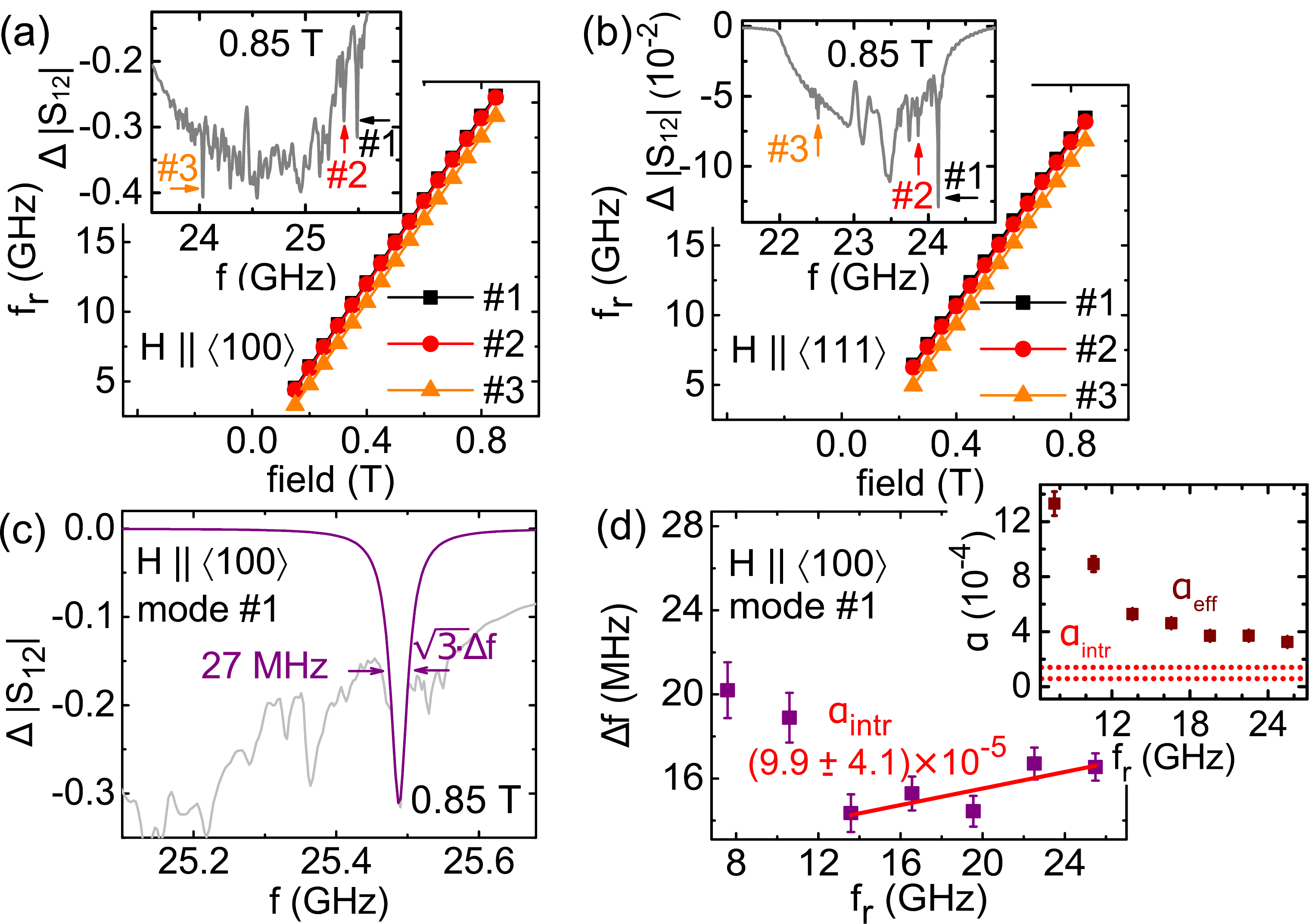}
	\caption{(Color online) (a)-(b) Resonance frequency as a function of field $H$ of selected sharp modes labelled \#1 to \#3 (see insets) for $\mathbf{H}\parallel\langle 100 \rangle$ and $\mathbf{H}\parallel\langle 111 \rangle$ at T = 5\,K. (c) Exemplary Lorentz fit of sharp mode \#1 for $\mathbf{H}\parallel\langle 100 \rangle$ at 0.85\,T. (d) Extracted linewidth $\Delta$f as a function of resonance frequency $f_{\text{r}}$ along with the linear fit performed to determine the intrinsic damping $\alpha_{\text{intr}}$ in \cso. Inset: Comparison among the extrinsic and intrinsic damping contribution. The red dotted lines mark the error margins of $\alpha_{\text{intr}}=(9.9\,\pm\,4.1)\times10^{-5}$.}
	\label{sharp_modes}
\end{figure}

The development of future devices for microwave applications, spintronics and magnonics \cite{2011::Zutic::NatMat,KrawcykGrundler2014,Chumak2015Review} requires materials with a low spin wave (magnon) damping. Insulating compounds are advantageous over metals for high-frequency applications as they avoid damping via spin wave scattering at free charge carriers and eddy currents \cite{Gurevich,Sparks}. Indeed, the ferrimagnetic insulator yttrium iron garnet (YIG) holds the benchmark with a Gilbert damping parameter $\alpha_{\text{intr}}=3\times10^{-5}$ at room temperature \cite{Serga2010,Klingler2017}. During the last years chiral magnets have attracted a lot of attention in fundamental research and stimulated new concepts for information technology~\cite{2013:Fert:NatureNano,2013:Nagaosa:NatureNano}. This material class hosts non-collinear spin structures such as spin helices and Skyrmions below the critical temperature $T_{\rm c}$ and critical field \hc\ \cite{2009:Muhlbauer:Science, 2010:Yu:Nature, 2012:Seki:Science}. Additionally, Dzyaloshinskii-Moriya interaction (DMI) is present that induces both the Skyrmion lattice phase and nonreciprocal microwave characteristics \cite{Seki2016PRB}. Low damping magnets offering DMI would generate new prospects by particularly combining complex spin order with long-distance magnon transport in high-frequency applications and magnonics \cite{0953-8984-27-50-503001,Garst2017}. At low temperatures, they would further enrich the physics in magnon-photon cavities that call for materials with small $\alpha_{\text{intr}}$ to achieve high-cooperative magnon-to-photon coupling in the quantum limit \cite{2013::Huebl::PRL,2014::Tabuchi::PRL,2014::Zhang::PRL,2014::Goryachev::PRAppl}.
\\ \indent
In this work, we investigate the Gilbert damping in \cso, a prototypical insulator hosting Skyrmions \cite{1977:Kohn:JPhysSocJpn, 2010:Belesi:PhysRevB,2012:Adams:PhysRevLett, 2012:Seki:PhysRevB}. This material is a local-moment ferrimagnet with $T_{c} = 58$\,K and magnetoelectric coupling~\cite{2012:Seki:PhysRevB2} that gives rise to dichroism for microwaves~\cite{2013:Okamura:NatCommun, 2015::Okamura, 2015:MochizukiPhysRevLett.114.197203}. The magnetization dynamics in \cso\ has already been explored \cite{2012:Onose:PhysRevLett,Schwarze2015,Seki2016PRB}. A detailed investigation on the damping which is a key quality for magnonics and spintronics has not yet been presented however. To evaluate $\alpha_{\text{intr}}$ we explore the field polarized state (FP) where the two spin sublattices attain the ferrimagnetic arrangement\cite{2010:Belesi:PhysRevB}. Using spectra obtained by two different coplanar waveguides~(CPWs), we extract a minimum $\alpha_{\text{intr}}$=(9.9\,$\pm$\,4.1)$\times$10$^{-5}$ at 5\,K, i.e.~only about four times higher than in YIG. We resolve numerous sharp resonances in our spectra and attribute them to modes that are confined modes across the macroscopic sample and allowed for by the low damping. Our findings substantiate the relevance of insulating chiral magnets for future applications in magnonics and spintronics.\\ \indent
From single crystals of \cso\ we prepared two bar-shaped samples exhibiting different crystallographic orientations. The samples had lateral dimensions of $2.3\times0.4\times0.3\,\mathrm{mm}^{3}$. They were positioned on CPWs that provided us with a dynamic magnetic field $\mathbf{h}$ induced by a sinusoidal current applied to the signal surrounded by two ground lines. We used two different CPWs with either a broad~\cite{SWMW} or narrow signal line width of $w_{\text s}$\,=\,1\,mm or 20\,$\mu$m, respectively \cite{suppInfo}. The central long axis of the rectangular \cso\ rods was positioned on the central axis of the CPWs. The static magnetic field $\mathbf{H}$ was applied perpendicular to the substrate with $\mathbf{H}\parallel\langle 100 \rangle$ and $\mathbf{H}\parallel\langle 111 \rangle$ for sample S1 and S2, respectively. The direction of $H$ defined the $z$-direction. The dynamic field component $\mathbf{h}\perp \mathbf{H}$ provided the relevant torque for excitation. Components $\mathbf{h}\parallel\mathbf{H}$ did not induce precessional motion in the FP state of \cso. We recorded spectra by a vector network analyzer using the magnitude of the scattering parameter $S_{12}$. We subtracted a background spectrum recorded at 1\,T to enhance the signal-to-noise ratio (SNR) yielding the displayed $\Delta|S_{12}|$. In Ref.~\cite{Klingler2017}, Klingler {\em et al.} have investigated the damping of the insulating ferrimagnet YIG and found that Gilbert parameters $\alpha_{\text{intr}}$ evaluated from both the uniform precessional mode and standing spin waves confined in the macroscopic sample provided the same values.
For \cso\ we evaluated $\alpha$ in two ways\cite{2015::Wei::JPhysD}. When extracting the linewidth $\Delta H$ for different resonance frequencies $f_\text{r}$, the Gilbert damping parameter $\alpha_{\text{intr}}$ was assumed to vary according to \cite{1985::ApplPhys::BHeinrich,Kalarickal::JAPL::2006}
\begin{equation}
\mu_0 \gamma \cdot\Delta H = 4\pi \alpha_{\text{intr}} \cdot f_\text{r} + \mu_0 \gamma \cdot\Delta H_0,
\label{DeltaH}
\end{equation}
where $\gamma$ is the gyromagnetic factor and $\Delta H_0$ the contribution due to inhomogeneous broadening. Equation\,(\ref{DeltaH}) is valid when viscous Gilbert damping dominates over scattering within the magnetic subsystem \cite{Lenz::PRB::2006}. When performing frequency-swept measurements at different fields $H$, the obtained linewidth $\Delta f$ was considered to scale linearly with the resonance frequency as \cite{1968::JApplPhys::Patton}
\begin{equation}
\Delta f = 2\alpha_{\text{intr}}\cdot f_{\text{r}} + \Delta f_0
\label{Deltaf},
\end{equation}
with the inhomogeneous broadening $\Delta f_0$. The conversion from Eq.\,(\ref{DeltaH}) to Eq.\,(\ref{Deltaf}) is valid when $f_{\text{r}}$ scales linearly with $H$ and $\mathbf{H}$ is applied along a magnetic easy or hard axis of the material \cite{Kuanr,Farle}. In Fig.\,\ref{spectra} (a) to (d) we show spectra recorded in the FP state of the material using the two different CPWs. For the same applied field $H$ we observe peaks residing at higher frequency $f$ for $\mathbf{H}\parallel\langle 100 \rangle$ compared to $\mathbf{H}\parallel\langle 111 \rangle$. From the resonance frequencies, we extract the cubic magnetocrystalline anisotropy constant $K=(-0.6\pm0.1)\cdot10^3$\,J/m$^3$ for \cso\ \cite{suppInfo}. The magnetic anisotropy energy is found to be extremal for $\langle100\rangle$ and $\langle111\rangle$ reflecting easy and hard axes, respectively \cite{suppInfo}. The saturation magnetization of \cso\ amounted to $\mu_0M_{\rm s}=0.13$~T at 5~K\cite{2012:Adams:PhysRevLett}.
\\
\indent
 Figure \,\ref{spectra} summarizes spectra taken with two different CPWs on two different \cso\ crystals exhibiting different crystallographic orientation in the field $H$. For the narrow CPW [Fig.\,\ref{spectra} (a) and (c)], we observed a broad peak superimposed by a series of resonances that all shifted to higher frequencies with increasing $H$. The field dependence excluded them from being noise or artifacts of the setup. Their number and relative intensities varied from sample to sample and also upon remounting the same sample in the cryostat (not shown). They disappeared with increasing temperature $T$ but the broad peak remained. For the broad CPW [Fig.\,\ref{spectra} (b) and (d)], we measured pronounced peaks whose linewidths were significantly smaller compared to the broad peak detected with the narrow CPW. We resolved resonances below the large peaks [arrows in Fig.\,\ref{spectra} (b)] that shifted with $H$ and exhibited an almost field-independent frequency offset from the main peaks that we will discuss later. It is instructive to first follow the orthodox approach and analyze damping parameters from modes reflecting the excitation characteristics of the CPW \cite{Schwarze2015}. Second, we follow Ref.~\cite{Klingler2017} and analyze confined modes.
\\
\indent
Lorentz curves (blue) were fitted to the spectra recorded with the broad CPW to determine resonance frequencies and linewidths. Note that the corresponding linewidths were larger by a factor of $\sqrt{3}$ compared to the linewidth $\Delta f$ that is conventionally extracted from the imaginary part of the scattering parameters \cite{Stancil}. The extracted linewidths $\Delta f$ were found to follow linear fits based on Eq.\,(\ref{Deltaf}) at different temperatures (details are shown in Ref.~\cite{suppInfo}). In Fig.\,\ref{deltaH_narrowCPW} (a) we show a resonance curve that was obtained as a function of $H$ taken with the narrow CPW at 15\,GHz. The curve does not show sharp features as $H$ was varied in finite steps (symbols). The linewidth $\Delta H$ (symbols) is plotted in Fig.\,\ref{deltaH_narrowCPW} (b) for different resonance frequencies and temperatures. The data are well described by linear fits (lines) based on Eq.\,(\ref{DeltaH}). Note that the resonance peaks measured with the broad CPW were extremely sharp. The sharpness did not allow us to analyze the resonances as a function of $H$. We refrained from fitting the broad peaks of Fig.\,\ref{spectra} (a) and (c) (narrow CPW) as they showed a clear asymmetry attributed to the overlap of subresonances at finite wavevector $k$, as will be discussed below.
\begin{figure}[h!t!]
	\includegraphics[width=1.0\linewidth]{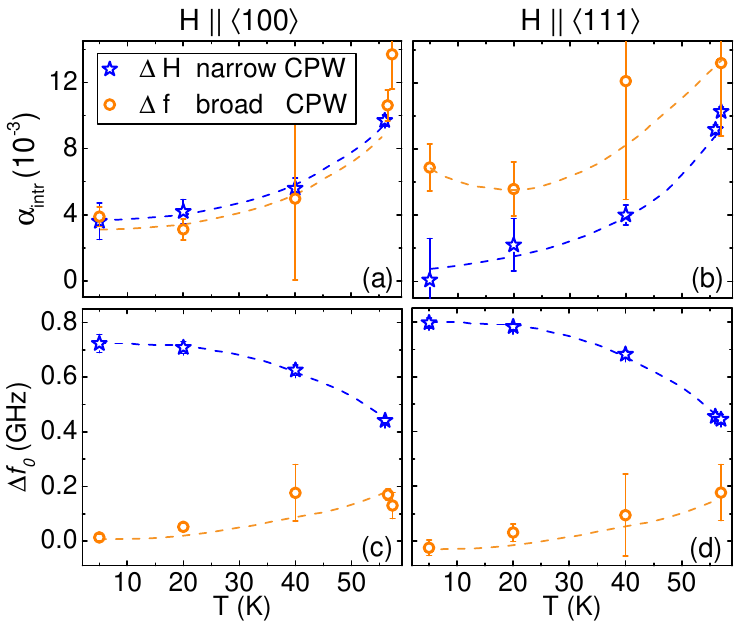}
	\caption{(Color online) (a) and (b) Intrinsic damping parameters $\alpha_{\text{intr}}$ and inhomogeneous broadening $\Delta f_0$ for two different field directions (see labels) obtained from the slopes and intercepts at $f_{\text{r}}=0$ of linear fits to the linewidth data (see Fig.\,\ref{deltaH_narrowCPW} (b) and Ref.\,\cite{suppInfo}). Dashed lines are guides to the eyes.}
	\label{damping_evaluation}
\end{figure}
\\ \indent
In Fig. \ref{damping_evaluation} (a) and (b) we compare the parameter $\alpha_{\text{intr}}$ obtained from both different CPWs (circles vs.~stars) and the two evaluation routes \footnote{We call it $\alpha_{\text{intr}}$ at this point as the parameter is extracted from linear slopes. Later we will show that standing spin waves provide the lowest $\alpha_{\text{intr}}$}. For $\mathbf{H}\parallel\langle 100 \rangle$ [Fig. \ref{damping_evaluation} (a)], between 5 and 20~K the lowest value for $\alpha_{\text{intr}}$ amounts to (3.7\,$\pm$\,0.4)$\times$10$^{-3}$. This value is three times lower compared to preliminary data presented in Ref.\,\cite{Schwarze2015}. Beyond 20\,K the damping is found to increase. For $\mathbf{H}\parallel\langle 111 \rangle$ [Fig.\,\ref{damping_evaluation} (b)] we extract (0.6\,$\pm$\,0.6)$\times$10$^{-3}$ as the smallest value. Note that these values for $\alpha_{\rm intr}$ still contain an extrinsic contribution and thus represent upper bounds for \cso, as we will show later. For the inhomogeneous broadening $\Delta f_0$ in Fig.\,\ref{damping_evaluation} (c) and (d) the datasets are consistent (we have used the relation $\Delta f_0=\gamma\Delta H_0 / 2\pi$ to convert $\Delta H_0$ into $\Delta f_0$). We see that $\Delta f_0$ increases with $T$ and is small for the broad CPW, independent of the crystallographic direction of $H$. For the narrow CPW the inhomogeneous broadening is largest at small $T$ and then decreases by about 40~\% up to about 50 K. Note that a CPW broader than the sample is assumed to excite homogeneously at $f_{\rm FMR}$ \cite{Iguchi::PRB::2015} transferring a wave vector $k=0$ to the sample. Accordingly we ascribe the intense resonances of Fig.\,\ref{spectra} (b) and (d) to $f_{\rm FMR}$. Using $f_{\rm FMR}$\,=\,6\,GHz and $\alpha_{\rm intr}$\,=\,$3.7\times 10^{-3}$ at 5 K [Fig.~\ref{damping_evaluation} (a)], we estimate a minimum relaxation time of $\tau=[2\pi\alpha_{\rm intr} f_{\rm r}]^{-1}=6.6\,$ns.
\\
\indent
In the following, we examine in detail the additional sharp resonances that we observed in spectra of Fig.\,\ref{spectra}. In Fig.\,\ref{spectra} (b) taken with the broad CPW for $\mathbf{H}\parallel\langle 100 \rangle$, we identify sharp resonances that exhibit a characteristic frequency offset $\delta f$ with the main resonance at all fields (black arrows). We illustrate this in Fig.\,\ref{Iguchi_vs_exp}(a) in that we shift spectra of Fig.\,\ref{spectra} (b) so that the positions of their main resonances overlap. The additional small resonances (arrows) in Fig.\,\ref{spectra} (b) are well below the uniform mode. This is characteristic for backward volume magnetostatic spin waves (BVMSWs). Standing waves of such kind can develop if they are reflected at least once at the bottom and top surfaces of the sample. The resulting standing waves exhibit a wave vector $k=n\pi/d$, with order number $n$ and sample thickness $d=0.3$\,mm. The BVMSW dispersion relation $f(k)$ of Ref.~\cite{Seki2016PRB} provides a group velocity $v_{\rm g}=-300$~km/s at $k=\pi/d$ [triangles in Fig.\,\ref{Iguchi_vs_exp} (b)]. Hence, the decay length $l_{\rm d}=v_{\rm g}\tau$ amounts to 2 mm considering $\tau=6.6\,$ns. This is larger than twice the relevant lateral sizes, thereby allowing standing spin wave modes to form in the sample. Based on the dispersion relation of Ref.~\cite{Seki2016PRB}, we calculated the frequency splitting $\delta f=f_{\rm FMR}-f(n\pi/d)$ [open diamonds in Fig.\,\ref{Iguchi_vs_exp} (b)] assuming $n=1$ and $t=0.4$\,mm for the sample width $t$ defined in Ref.~\cite{Seki2016PRB}. Experimental values (filled symbols) agree with the calculated ones (open symbols) within about 60 MHz. In case of the narrow CPW, we observe even more sharp resonances [Fig.\,\ref{spectra} (a) and (c)]. A set of resonances was reported previously in the field-polarized phase of \cso\ \cite{2010::Kobets::LTP,2012::Maisuradze::PPRL,2012:Onose:PhysRevLett,2015::Okamura}. Maisuradze {\em et al.}~assigned secondary peaks in thin plates of \cso\ to different standing spin-wave modes \cite{2012::Maisuradze::PPRL} in agreement with our analysis outlined above.
\\
\indent
The inhomogeneous dynamic field $\mathbf{h}$ of the narrow CPW provides a much broader distribution of $\mathbf{k}$ compared to the broad CPW. This is consistent with the fact that the inhomogeneous broadening $\Delta f_0$ is found to be larger for the narrow CPW compared to the broad one [Fig.\,\ref{damping_evaluation} (c) and (c)]. Under these circumstances, the excitation of more standing waves is expected. We attribute the series of sharp resonances in Fig.~\ref{spectra} (a) and (c) to such spin waves. In Fig.\,\ref{sharp_modes} (a) and (b) we highlight prominent and particularly narrow resonances with \#1, \#2 and \#3 recorded with the narrow CPW. We trace their frequencies $f_{\rm r}$ as a function of $H$ for $\mathbf{H}\parallel\langle 100 \rangle$ and $\mathbf{H}\parallel\langle 111 \rangle$, respectively. They depend linearly on $H$ suggesting a Land\'e factor $g=2.14$ at 5 K.\\ \indent We now concentrate on mode \#1 for $\mathbf{H}\parallel\langle 100 \rangle$ at 5~K that is best resolved. We fit a Lorentzian lineshape as shown in Fig.\,\ref{sharp_modes}(c) for 0.85\,T, and summarize the corresponding linewidths $\Delta f$ in Fig.\,\ref{sharp_modes}(d). The inset of Fig.\,\ref{sharp_modes}(d) shows the effective damping $\alpha_{\text{eff}}=\Delta f / (2f_{\text{r}})$ evaluated directly from the linewidth as suggested in Ref. \cite{Schwarze2015}. We find that $\alpha_{\text{eff}}$ approaches a value of about 3.5 $\times 10^{-4}$ with increasing frequency. This value includes both the intrinsic damping and inhomogeneous broadening but is already a factor of 10 smaller compared to $\alpha_{\text{intr}}$ extracted from Fig. \ref{damping_evaluation} (a). Note that \cso\ exhibiting 3.5 $\times 10^{-4}$ outperforms the best metallic thin-film magnet \cite{TomSilvaNeNatPhys}. To correct for inhomogeneous broadening and determine the intrinsic Gilbert-type damping, we apply a linear fit to the linewidths $\Delta f$ in Fig.\,\ref{sharp_modes}(d) at $f_{\text{r}}>10.6$\,GHz and obtain (9.9\,$\pm$\,4.1)$\times$10$^{-5}$. For $f_{\text{r}}\leq$ 10.6\,GHz the resonance amplitudes of mode \#1 were small reducing the confidence of the fitting procedure. Furthermore, at low frequencies, we expect anisotropy to modify the extracted damping, similar to the results in Ref.\,\cite{Silva1999}. For these reasons, the two points at low $f_{\text{r}}$ were left out for the linear fit providing (9.9\,$\pm$\,4.1)$\times$10$^{-5}$.
\\ \indent We find $\Delta f$ and the damping parameters of Fig.~\ref{damping_evaluation} to increase with $T$. It does not scale linearly for $\mathbf{H}\parallel\langle 100 \rangle$ \cite{suppInfo}. A deviation from linear scaling was reported for YIG single crystals as well and accounted for by the confluence of a low-$k$ magnon with a phonon or thermally excited magnon \cite{Sparks}. In the case of $\mathbf{H}\parallel\langle 111 \rangle$ (cf.\,Fig.\,\ref{damping_evaluation} (b)) we obtain a clear discrepancy between results from the two evaluation routes and CPWs used. We relate this observation to a misalignment of $\mathbf{H}$ with the hard axis $\langle  111 \rangle$. The misalignment motivates a field-dragging contribution \cite{Farle} that can explain the discrepancy. For this reason, we concentrated our standing wave analysis on the case $\mathbf{H}\parallel\langle 100 \rangle$. We now comment on our spectra taken with the broad CPW that do not show the very small linewidth attributed to the confined spin waves. The sharp mode \#1 yields $\Delta f=15.3$\,MHz near 16\,GHz [Fig.\,\ref{sharp_modes} (d)]. At 5\,K the dominant peak measured at 0.55\,T with the broad CPW provides however $\Delta f=129$~MHz. $\Delta f$ obtained by the broad CPW is thus increased by a factor of eight and explains the relatively large Gilbert damping parameter in Fig.\,\ref{damping_evaluation} (a) and (b). We confirmed this larger value on a third sample with $\mathbf{H}\parallel\langle 100 \rangle$ and obtained (3.1\,$\pm$\,0.3)$\times$10$^{-3}$ \cite{suppInfo} using the broad CPW. The discrepancy with the damping parameter extracted from the sharp modes of Fig.\,\ref{sharp_modes} might be due to the remaining inhomogeneity of $\mathbf{h}$ over the thickness of the sample leading to an uncertainty in the wave vector in $z$-direction. For a standing spin wave such an inhomogeneity does not play a role as the boundary conditions discretize $k$. Accordingly, Klingler {\em et al.} extract the smallest damping parameter of $2.7(5)\times10^{-5}$ reported so far for the ferrimagnet YIG when analyzing confined magnetostatic modes \cite{Klingler2017}.\\ \indent
To summarize, we investigated the spin dynamics in the field-polarized phase of the insulating chiral magnet \cso. We detected numerous sharp resonances that we attribute to standing spin waves. Their effective damping parameter is small and amounts to $3.5\times 10^{-4}$. A quantitative estimate of the intrinsic Gilbert damping parameter extracted from the confined modes provides even $\alpha_{\text{intr}}$=(9.9\,$\pm$\,4.1)$\times$10$^{-5}$ at 5\,K. The small damping makes an insulating ferrimagnet exhibiting Dzyaloshinskii-Moriya interaction a promising candidate for exploitation of complex spin structures and related nonreciprocity in magnonics and spintronics.\\ \indent We thank S.\ Mayr for assistance with sample preparation. Financial support through DFG TRR80, DFG 1143, DFG FOR960, and ERC Advanced Grant 291079 (TOPFIT) is gratefully acknowledged.

\bibliography{CuOSeO_damping}

\end{document}